# Engineering of intrinsic chiral torques in magnetic thin films based on the Dzyaloshinskii-Moriya interaction

Zhentao Liu[1,2], Zhaochu Luo[1,2], Stanislas Rohart[3], Laura J. Heyderman[1,2], Pietro Gambardella[4], Aleš Hrabec[1,2,4]

[1]Laboratory for Mesoscopic Systems, Department of Materials, ETH Zurich, 8093 Zurich, Switzerland

[2]Laboratory for Multiscale Materials Experiments, Paul Scherrer Institute, 5232 Villigen PSI, Switzerland

[3]Université Paris-Saclay, CNRS, Laboratoire de Physique des Solides, 91405, Orsay, France

[4]Laboratory for Magnetism and Interface Physics, Department of Materials, ETH Zurich, 8093 Zurich, Switzerland

Authors to whom correspondence should be addressed: ales.hrabec@psi.ch; zhentao.liu@psi.ch

**Abstract:**

The establishment of chiral coupling in thin magnetic films with inhomogeneous anisotropy has led to the development of artificial systems of fundamental and technological interest. The chiral coupling itself is enabled by the Dzyaloshinskii-Moriya interaction (DMI) enforced by the patterned non-collinear magnetization. Here, we create a domain wall track with out-of-plane magnetization coupled on each side to a narrow parallel strips with in-plane magnetization. With this we show that the chiral torques emerging from the DMI at the boundary between the regions of noncollinear magnetization in a single magnetic layer can be used to bias the domain wall velocity. To tune the chiral torques, the design of the magnetic racetracks can be modified by varying the width of the tracks or the width of the transition region between noncollinear magnetizations, reaching effective chiral magnetic fields of up to 7.8 mT. Furthermore, we show how the magnitude of the chiral torques can be estimated by measuring asymmetric domain wall velocities, and demonstrate spontaneous domain wall motion propelled by intrinsic torques even in the absence of any external driving force.

**Main Text:**

The engineering of magnetic torques in magnetic nanostructures is central to the development of spintronic devices for nonvolatile information storage [1-6], Boolean logic [7-10] and neuro-inspired computation [11-13]. Magnetic torques can be induced by electric currents [1,3] or by coupling to a nearby magnetic layer, for example via the Ruderman-Kittel-Kasuya-Yosida interaction [6,14], the exchange interaction [15,16] and the dipolar interaction [17]. The implementation of interlayer coupling effects into magnetic devices has led to enhanced device performance such as improved thermal stability [18-20], lower power consumption and higher operating speed [21-25]. However, magnetic torques resulting from interlayer coupling offer limited possibilities for the operation of magnetic elements lying in the same plane, such as those found in planar magnetic logic circuits [9,26] and artificial spin ice systems [27]. The possibility to induce intralayer coupling between adjacent nanomagnets offered by the interfacial DMI therefore opens alternative perspectives in this respect. Indeed, in recent work performed on DMI-coupled systems, a number of phenomena arising from the lateral coupling have been demonstrated, such as lateral exchange bias between adjacent regions of the same magnetic layer, two-dimensional synthetic antiferromagnets, and field-free switching of coupled nanomagnets by spin-orbit torques [28]. Moreover, the intralayer DMI can be used to nucleate and define the position of magnetic domain walls (DW) [29,30] and realize all-electrical DW logic circuits [10,26,31,32].

Here we show that the intralayer DMI can be used universally to engineer strong, localized intrinsic chiral torques that trigger the spontaneous motion of DWs or bias the speed of current-driven DWs in magnetic racetracks. We first establish a versatile method to characterize the intralayer DMI, which is based on asymmetric DW motion in a benchmark Pt/Co/AlOx trilayer. We then further extend the method to a Pt/CoB/AlOx trilayer with low DW pinning, where the intrinsic chiral torque due to the DMI is utilized to move a DW in the absence of any external force.

The intralayer chiral coupling induced by the interfacial DMI occurs between adjacent regions of the same magnetic layer having in-plane (IP) and out-of-plane (OOP) anisotropy. These regions can be patterned by selective oxidation or ion irradiation [28,33]. As in bulk systems, the DMI favors the orthogonal alignment of neighboring magnetic moments. The simplest IP-OOP element, schematically shown in Fig. 1(a), can attain four possible magnetic configurations: ←⊙, ←⊗, →⊙ and →⊗, where ← and → (⊙ and ⊗) represent the orientation of the magnetization in the IP (OOP) regions. Due to the presence of DMI in Pt/Co/AlOx [10,28,34,35], the left-handed ←⊗ and →⊙ configurations have lower energy, and are separated from the ←⊙ and →⊗ high energy states by the energy barrier given by shape anisotropy in the IP element or by the uniaxial anisotropy in the OOP element. To a first approximation, the energy difference between these states is determined by the DMI energy per unit length of a $\pi/2$ domain wall, $E_{\mathrm{DM}} = \pm \pi D t/2$ [36], where $D$ is the DMI constant and $t$ is the thickness of the ferromagnet. The chiral alignment of IP ($\vec{M}_{\mathrm{IP}}$) and OOP ($\vec{M}_{\mathrm{OOP}}$) magnetic moments is thus due to a local effective DMI field [29,37]

$$\vec{H}_{\mathrm{DM}} = \frac{2D}{\mu_0 M_s}\left(-\frac{\partial m_z}{\partial x}, 0, \frac{\partial m_x}{\partial x}\right), \qquad (1)$$

where $\mu_0$ is the vacuum permeability and $M_s$ is the saturation magnetization. $m_z$ and $m_x$ are the z and x components of the normalized magnetization vector with the coordinates defined in Fig. 1(a). In order to observe the spontaneous effect of chiral coupling in such basic elements, the anisotropy has to be low-enough so that the effective magnetic field can overcome the associated energy barrier. This approach has been used to form spontaneous chiral magnetic textures such as skyrmions and polar vortices, as well as DW injectors, inverters or logic devices [10,26,29,31,32].

To determine the effect of the chiral coupling at OOP-IP boundary, we propose a method based on asymmetric DW motion. A DW racetrack device is defined in an extended film as a central track with OOP magnetization, surrounded by two narrow parallel IP strips, and extended OOP regions which have a uniform magnetization set by external magnetic field, as shown in Fig. 1(b). This defines two stable states, where the magnetization in the central track is parallel ($\odot\odot\odot$) or antiparallel ($\odot\otimes\odot$) to the two external OOP regions. By injecting a DW into the central track, the energy of the system varies with the DW potion along the track, as schematically shown in Fig. 1(b), since the chiral coupling across the IP strips favors a track magnetization antiparallel to that of the OOP extended regions (i.e. $\odot\otimes\odot$ state). This situation can be represented (see Appendix for the detailed model) by an out-of-plane effective driving field:

$$H_{\text{eff}} = \frac{E_{\odot\otimes\odot} - E_{\odot\odot\odot}}{2\,t\,\mu_0 M_s w_{\text{OOP}}}, \qquad (2)$$

where $E_{\odot\otimes\odot}$ and $E_{\odot\odot\odot}$ are the linear energy densities of the favorable ($\odot\otimes\odot$) and unfavorable ($\odot\odot\odot$) magnetic states, $w_{\text{OOP}}$ is the width of the central track and $t$ is the thickness of the ferromagnetic layer. Since $E_{\odot\otimes\odot} < E_{\odot\odot\odot}$, $H_{\text{eff}}$ induces a DW motion along +*x* as shown in Fig. 1(b). This effective field is thus due to the energy gradient depending on the DW position along the track. Moreover, in addition to the chiral coupling at the track edges, the energy difference between $E_{\odot\odot\odot}$ and $E_{\odot\otimes\odot}$ states also results from the dipolar energy (flux closure between the extended OOP regions and the track). The two effects can be distinguished based on their different dependence on the track width. The chiral coupling at the track edges is a local effect that does not depend on the track width, and therefore $H_{\text{eff}}$ scales as $1/w_{\text{OOP}}$. In contrast, the dipolar coupling increases with the track width, and therefore $H_{\text{eff}}$ should decrease slower than $1/w_{\text{OOP}}$ [38]. In the structures investigated here, we find that, depending on the magnitude of the DMI and the thickness, the effects play different roles: for Pt/Co/AlOx, the DMI is the dominating effect, whereas both effects contribute substantially in Pt/CoB/AlOx trilayers.

We experimentally investigate this situation in Pt/Co/Al and Pt/CoB/Al based samples, where the anisotropy can be tuned by a local Al oxidation. We used magnetron sputtering and electron beam lithography to deposit and pattern a Pt(5nm)|Co(1.6nm)|Al(2nm) multilayer (see Supplementary materials [38] for the fabrication details). The regions with OOP and IP magnetic anisotropy were defined by selective oxidation of the top Al layer through a mask using an oxygen plasma, which protects and defines the regions with IP magnetization. The width of the central track (see Fig. 1(b)) $w_{OOP}$ ranges from 350 nm to 950 nm, and the width of the IP strips $w_{IP}$ is kept constant at 55 nm.

In order to reproducibly inject DWs into the track, we create an additional IP V-shaped at both ends of the central track [see Fig. 1(c)]. The V-shape structure facilitates the nucleation of a reversed domain upon current injection [10,31]. To investigate the propagation of DWs in the central track, we first apply a large external OOP magnetic field (+/- 75 mT) to set the magnetic regions in the central DW track and surrounding IP strips to either ⊙ (with a $+H_z$ field) or ⊗ (with a $-H_z$ field), resulting in an unfavorable configuration. This is followed by a sequence of current pulses to nucleate the DWs at the ends of the central racetrack. Then, in order to move the DWs, we apply a series of 50-ns-long current pulses separated by 100 ms. The current density is varied from $1.5 \times 10^{11}$ A/m$^2$ to $4 \times 10^{11}$ A/m$^2$, and the current polarity determines the direction of motion of the DWs, which are of Néel type [36,39].

The position of the DW is tracked by polar Kerr microscopy and a set of images of the DW motion in a racetrack with $w_{OOP}$ = 950 nm at a current density of $3.6 \times 10^{11}$ A/m$^2$ is shown in Fig. 1(c) after applying a series of current pulses along the +x direction. The DW motion induced by the spin-orbit torques is expected to be the same for both ⊙|⊗ and ⊗|⊙ DWs (up-down and down-up mobile DWs); in other words, both the magnitude and sign of the DW displacement for a given polarity of the current should be equal in the absence of external fields [36,40,41]. However, a clear asymmetry is observed in Fig. 1(c) since only the ⊗|⊙ DW (⊙|⊗ DW) moves with a positive current and initialization with a with a $+H_z$ field ($-H_z$ field). This is a result of the fact that the DW that moves favors the expansion of the magnetic domain that satisfies the DMI-induced antiferromagnetic alignment between the central OOP track and the outer OOP strips. The other DW remains fixed at this current density, because its motion would lead to the expansion of a domain with magnetization parallel to that of the outer OOP region, which is not favorable. We thus conclude that the DMI coupling at the boundaries of the racetrack exerts an effective chiral torque on the DWs, which is the cause of their asymmetric motion.

To probe the dependence of the chiral coupling on the width of the racetrack, we performed a series of measurements of the DW velocity on the current density in tracks with different $w_{OOP}$, as shown in Fig. 2(a). It can be seen that the DW velocity is strongly asymmetric, and the velocity of the favored DW increases upon narrowing $w_{OOP}$, as expected due to the increased importance of the edge of the narrower tracks. In addition, we observe no asymmetry in a control experiment performed on a 3-µm-wide racetrack without the IP magnetized strips ($w_{IP}$ = 0). Moreover, because the chiral coupling acts as an effective magnetic field $H_{eff}$ given by Eq. (2), the observed asymmetry can be compensated by applying an external magnetic field. As shown in Fig. 2(b), on applying an OOP magnetic field $H_z$ ranging from 0 mT to ±10 mT in a track with $w_{OOP}$ =350 nm, the motion of the mobile DW is gradually suppressed, whereas the velocity of the unfavorable DW increases. The strength of $H_{eff}$ can in principle be estimated to be the value of $H_z$ that gives no difference in DW velocity for positive and negative current. However, this field cannot be reached experimentally due to parasitic domain nucleation at higher fields.

To quantify $H_{\text{eff}}$, we therefore measured the DW motion due to the magnetic field $H_z$ alone, i.e. without applying an electric current. In Fig. 3(a), measurements of all track widths are shown without and with chiral coupling ($w_{\text{IP}}$ = 0 and 55 nm) in the low-field creep regime on applying $H_z$ in pulses of duration between 2 s and 50 s. The DW velocity in the track without chiral coupling is symmetric with respect to the field direction, whereas it is asymmetric in the tracks with the IP strips. The asymmetry increases upon increasing the chiral coupling strength by narrowing the track width $w_{\text{OOP}}$. The DW velocity at low magnetic fields generally follows the thermally activated domain wall creep law [42], which can be expressed as

$$v = v_0 \exp\left[-\xi\big(\mu_0 H_z \pm \mu_0 H_{\text{eff}}(w_{\text{OOP}})\big)^{-1/4}\right], \quad (3)$$

where $v_0$ is the characteristic speed and $\xi$ is the scaling coefficient proportional to the thermal energy. The effect of chiral coupling can then be taken into account by adding the DMI effective field $H_{\text{eff}}(w_{\text{OOP}})$ to the external field. By fitting the DW velocity as a function of $H_z$, we thus find the effective chiral coupling field $H_{\text{eff}}$ at different values of $w_{\text{OOP}}$, as shown in Fig. 3(b). The maximum effective DMI field of 7.8 mT corresponds to the minimum $w_{\text{OOP}}$ = 350 nm. As expected from Eq. (2), we confirm that $H_{\text{eff}}$ is inversely proportional to $w_{\text{OOP}}$, as dictated by the local origin of the effective DMI field [28].

Although the chiral torque due to $H_{\text{eff}}$ has a strong impact on the DW propagation, the torque is not strong enough to induce spontaneous DW motion due to the strong pinning of DWs in Pt/Co/AlOx. Indeed, it is well-known that disorder hinders the DW dynamics by pinning the DWs in Pt/Co films [43-45] and therefore the energy landscape cannot be simply described by Eq. (2) but rather resembles the one sketched in Fig. 1(b). We therefore turn to an alternative material system based on Pt(5nm)|CoB(4.8nm)|Al(2nm), which is known for its low pinning [46-48] and apply the same methodology. IP and OOP magnetized regions are again defined by selective oxidation of the Al capping layer, which changes the magnetic anisotropy of CoB from IP to OOP, as shown in the Supplementary materials [38,49]. A similar track design is used to demonstrate the relationship between $H_{\text{eff}}$ and $w_{\text{OOP}}$ as well as $w_{\text{IP}}$. To examine the effect of the chiral torque due to $H_{\text{eff}}$, the magnetic state of the entire layer is first initialized by applying a magnetic field $H_z = \pm 75$ mT. This results in a high-energy configuration as before, in which the OOP racetrack and the outer OOP region have an ⊙ (⊗) magnetization. As a result of the low pinning and sufficient chiral coupling strength, the DWs start moving before the magnetic field is switched off. Thus, instead of measuring the DW velocity as a function of magnetic field, we measured the so-called stopping field, i.e., the field $H_z$ that compensates $H_{\text{eff}}$ (see supplementary materials [38,50,51] for details). The stopping field is measured by decreasing $H_z$ to zero after saturation in steps of 0.1 mT and observing the threshold value of $H_z$ at which the DW motion is arrested. This field has been measured for devices with $w_{\text{OOP}}$ ranging from 200 nm to 450 nm with $w_{\text{IP}}$ = 50 nm. As shown in Fig. 4(a), the stopping field monotonically increases upon reducing $w_{OOP}$, up to ±4.2 mT for $w_{OOP}$ = 200 nm. This observation is in qualitative agreement with the proposed model in Eq. (2). Interestingly, the combination of a low pinning and large chiral coupling effect leads to a spontaneous DW motion at zero magnetic field.

To investigate the dependence of $H_{\text{eff}}$ on the size of the IP magnetized strips, the stopping field was measured in devices with $w_{IP}$ ranging from 30 nm to 90 nm with a fixed $w_{OOP}$ = 200 nm [see Fig. 4(b)]. In contrast to the case of variable $w_{OOP}$, the dependence of the stopping field is not monotonous and a maximum field of ±4.4 mT is observed for $w_{IP}$ = 60 nm. We can understand this by considering that, for $w_{IP}$ narrower than 60 nm, the size of the IP track approaches the natural DW width given by anisotropy and exchange coupling constant, which suppress the chiral coupling strength. For $w_{IP} \gtrsim 60$ nm, the coupling is gradually lost due to the fact that the chiral coupling effect is localized at the OOP|IP interface. To corroborate our findings, we have performed micromagnetic simulations of DW propagation in Pt/CoB/AlOx racetracks using MuMax3 (see Supplementary materials [38] for details of the simulation parameters). The simulations, performed in the absence of pinning, reproduce the spontaneous motion of DWs in the racetracks that results from the chiral coupling between the magnetization of the inner track and the outer OOP region [see Fig. 4(c)], and a $1/w_{OOP}$ dependence of stopping field is observed while changing the width of OOP region with DMI constants of 0.1 mJ/m$^2$ and 0.2 mJ/m$^2$. Both experimental [Fig. 4(a)] and simulation data [Fig. 4(c)] show qualitatively the same $1/w_{OOP}$ dependence. In addition, as highlighted in Fig. 4(d), the maximum stopping field for a given $w_{IP}$ can be reproduced. However, the position of the maximum strongly depends on the amount of the OOP anisotropy present in the IP strip. Such an anisotropy contribution is expected since, despite the magnetization lying in plane, there is still a significant contribution to the anisotropy from the Pt/CoB interface.

To provide more quantitative insight into the magnitude of $H_{\text{eff}}$, the energies $E_{\odot\odot\odot}$ and $E_{\odot\otimes\odot}$ are determined from 1D micromagnetic simulations. We consider a varying magnetic texture across a track with periodic boundary conditions along *x* and where we vary $w_{OOP}$ and $w_{IP}$. For simplicity, the OOP anisotropy in the IP region is set to zero. As can be seen in Fig. 5(a), $H_{\text{eff}}$ does not completely follow the $1/w_{OOP}$ trend for $w_{OOP} > 50$ nm due to the fact that the dipolar energy also contributes to the effective field (see Supplementary materials [38] for the estimation). We also confirm that $H_{\text{eff}}$ attains a maximum value at $w_{IP} \approx 20$ nm for different $w_{OOP}$ [Fig 5(b), in good agreement with the stopping field calculations. For narrower (wider) IP widths $w_{IP}$ the magnitude of $H_{\text{eff}}$ drops off because of the same reasons as described above. This therefore suggests that the width of the IP region has to be carefully tailored for a given application. A full dependence of $H_{\text{eff}}$ on $w_{OOP}$ and $w_{IP}$ in Fig. 5(c) complements the results shown in Fig. 5(a-b).

In conclusion, we have shown that the intrinsic chiral torques arising from the DMI at the boundary between the OOP and IP regions of Pt/Co/AlOx and Pt/CoB/AlOx DW tracks result in strongly asymmetric DW motion affecting both the current- and field-induced dynamics. The effective field $H_{\text{eff}}$ corresponding to the chiral torques can be quantified by measuring the DW velocity in the creep regime or the field required to stop the spontaneous motion of DWs driven by the DMI. Our measurements show that $H_{\text{eff}}$ can be finely tuned by designing racetracks with different lateral dimensions $w_{\text{OOP}}$ and $w_{\text{IP}}$. The observation of strong chiral torques in materials with low pinning, such as CoB, makes it possible to manipulate DWs in the absence of external magnetic fields or electric currents. The unidirectional domain wall motion and the spontaneous domain wall propagation could be used to accelerate or decelerate DWs in selected regions of a racetrack when driven by a constant electric current as well as to implement the integrate-and-fire process in artificial neuron network applications. These findings open an alternative path to the local engineering of magnetic torques, which could be implemented for DW motion-based neuromorphic devices [52-55] and magnonic devices [56-59].

**Acknowledgement:**
This project received funding from the Swiss National Science Foundation (Grant Agreements No. 182013 and 200020_200465). A.H. was funded by the European Union's Horizon 2020 research and innovation program under Marie Skłodowska-Curie grant agreement No. 794207 (ASIQS). The authors thank Aleksandr Kurenkov for technical support.

**Data Availability:**
The data that support this study are available via the Zenodo repository, https://doi.org/10.5281/zenodo.5045245, Ref. [60]

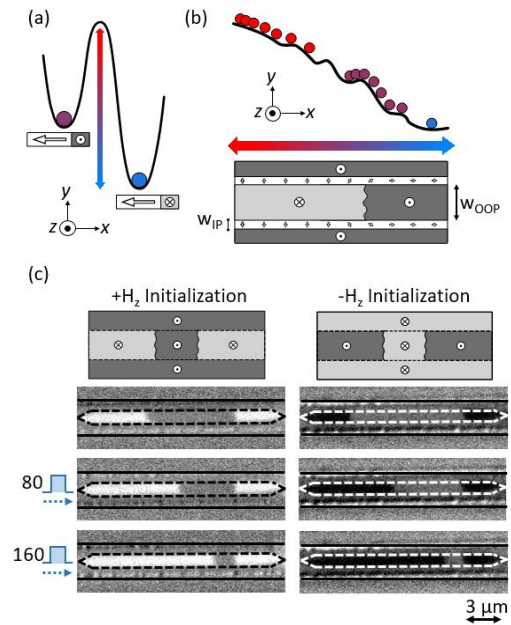

**FIG. 1**. (a) Schematic of the energy barrier between the unfavorable ←⊙ and favorable ←⊗ chiral configurations of adjacent IP-OOP magnetic regions. The energy barrier between the favourable and unfavourable configuration is given mainly by the shape and uniaxial anisotropies of the IP and OOP regions, respectively. (b) Schematic of a DW racetrack with OOP magnetization delimited by two strips with IP anisotropy. The width of the OOP track and IP strips are $w_{OOP}$ and $w_{IP}$, respectively. The energy landscape of the DW racetrack is monotonous, except for local DW pinning sites, with the slope given by the energy difference between $E_{\odot\odot\odot}$ and $E_{\odot\otimes\odot}$. (c) Sequence of Kerr images of asymmetric DW motion in a racetrack with $w_{OOP}$ = 950 nm with corresponding schematics. The dashed lines correspond to the position of the IP magnetized strip whereas the full lines correspond to the boundary of the magnetic track. 80 current pulses are applied between each image; the pulse duration is 50 ns and the current density $3.6 \times 10^{11}\ A/m^2$. The initial magnetic configuration of both inner and outer OOP regions was ⊙ and ⊗ in the left and right panel, respectively.

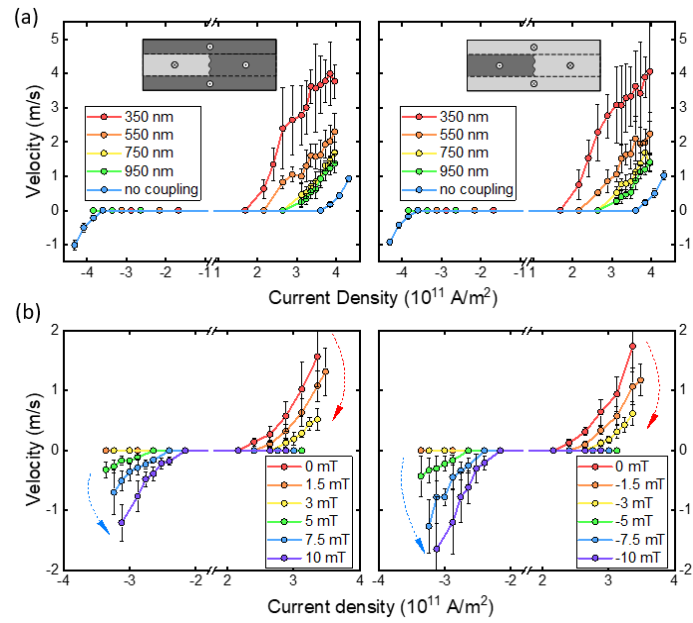

**FIG. 2**. (a) DW velocity as function of current density for DW tracks with fixed $w_{IP} = 55$ nm and $w_{OOP}$ ranging from 350 nm to 950 nm with $+H_z$ initialization (left panel) and $-H_z$ initialization (right panel) . The lines are guides to the eye. The error bars are standard deviations for 10 individual measurements. (b) Current induced DW velocity for different values of the OOP magnetic field $H_z$. The red and blue arrows show the trend of suppressing and enhancing the DW velocity with increasing field, respectively.

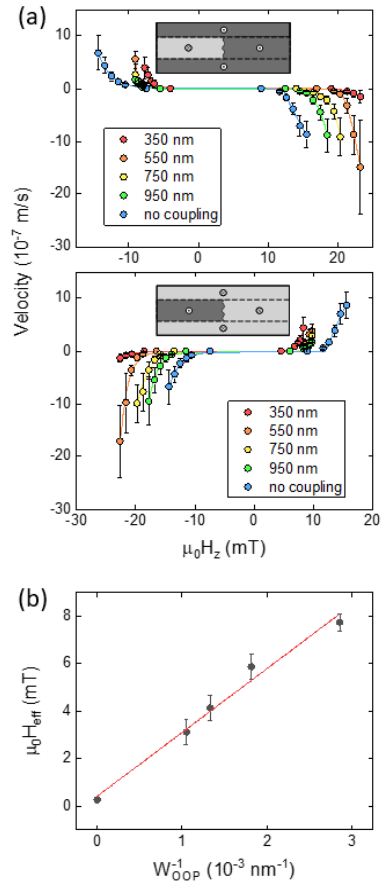

**FIG. 3**. (a) DW velocity induced by an OOP field $H_z$ measured in Pt/Co/AlOx devices with $w_{IP} = 55\ nm$ and for various $w_{OOP}$ with $+H_z$ initialization (upper panel) and $-H_z$ initialization (lower panel). Control measurements of a 3 μm-wide DW track without the IP strips are also shown. The lines are fits according to Eq. (3). (b) Effective chiral coupling strength ($H_{eff}$) derived from the creep model. The red line is a linear fitting respect to $1/w_{OOP}$. Error bars represent the standard deviation of $H_{eff}$ measured using both positive and negative magnetic fields.

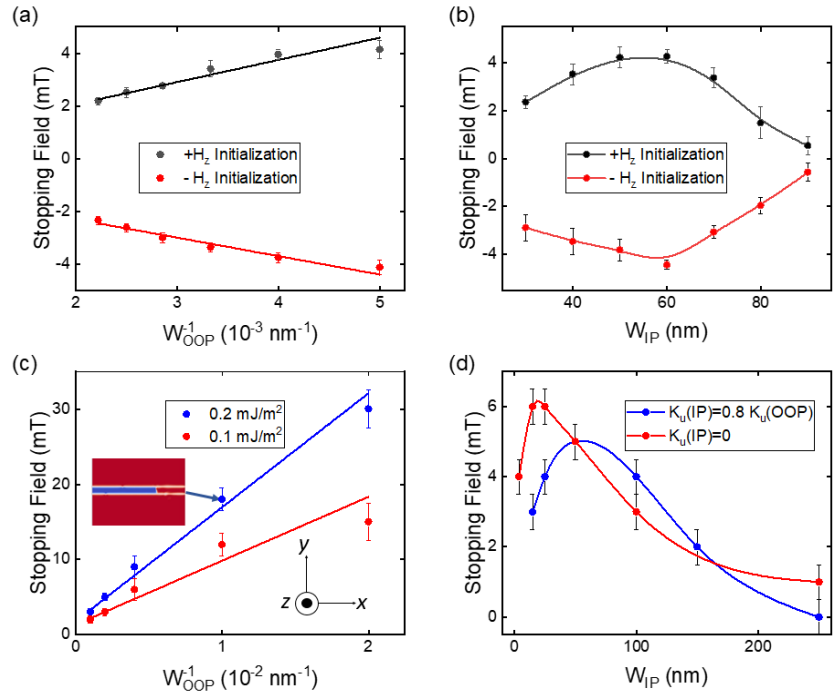

**FIG. 4**. (a) Stopping field versus $1/w_{OOP}$ for the DWs in Pt/CoB/AlOx racetracks with $w_{IP}$ = 50 nm. (b) Stopping field in racetracks with $w_{OOP}$ = 200 nm and $w_{IP}$ ranging from 30 nm to 90 nm. (c-d) Micromagnetic simulations of the stopping field in a track with a fixed $w_{IP}$ = 50 nm and variable $1/w_{OOP}$ with different interfacial DMI strengths $D$=0.1 and 0.2 mJ/m$^2$ (c) and in a track with fixed $w_{OOP}$ = 200 nm and variable $w_{IP}$ (d) where $D$ = 0.2 mJ/m$^2$ with different uniaxial anisotropy constants ($K_u$). The inset in (c) shows a snapshot of the simulations. Red (blue) color correspond to the OOP magnetization pointing along the +z (-z) direction. The lines in (a) and (c) are linear fits of the stopping field vs $1/w_{OOP}$ whereas the lines in (b) and (d) are guides to the eye.

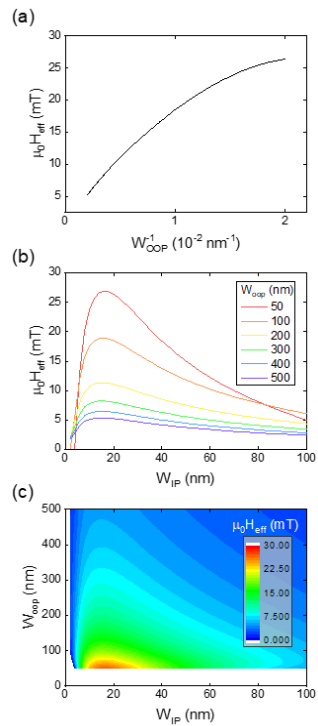

**FIG. 5**. (a) 1D micromagnetic calculation of $H_{eff}$ versus $1/w_{OOP}$ in a Pt/CoB/AlOx racetrack with fixed $w_{IP}$ = 20 nm. $H_{eff}$ is extracted from the calculation of the linear energy density of ⊙⊙⊙ and ⊙⊗⊙ regions using Eq. (2) with $D$ = 0.2 mJ/m². (b) Simulated $H_{eff}$ as a function of $w_{IP}$ for $w_{OOP}$ = 50 to 500 nm. (c) Color map of $H_{DM}^z$ versus $w_{IP}$ and $w_{OOP}$.

**Reference:**

[1] A. Brataas, A. D. Kent, and H. Ohno, Current-induced torques in magnetic materials, Nat. Mater. **11**, 372 (2012).

[2] G. Fuchs, N. Emley, I. Krivorotov, P. Braganca, E. Ryan, S. Kiselev, J. Sankey, D. Ralph, R. Buhrman, and J. Katine, Spin-transfer effects in nanoscale magnetic tunnel junctions, Appl. Phys. Lett. **85**, 1205 (2004).

[3] A. Manchon, J. Železný, I. M. Miron, T. Jungwirth, J. Sinova, A. Thiaville, K. Garello, and P. Gambardella, Current-induced spin-orbit torques in ferromagnetic and antiferromagnetic systems, Rev. Mod. Phys. **91**, 035004 (2019).

[4] E. Grimaldi, V. Krizakova, G. Sala, F. Yasin, S. Couet, G. S. Kar, K. Garello, and P. Gambardella, Single-shot dynamics of spin–orbit torque and spin transfer torque switching in three-terminal magnetic tunnel junctions, Nat. Nanotech. **15**, 111 (2020).

[5] S. S. Parkin, M. Hayashi, and L. Thomas, Magnetic domain-wall racetrack memory, Science **320**, 190 (2008).

[6] P. Grünberg, R. Schreiber, Y. Pang, M. Brodsky, and H. Sowers, Layered magnetic structures: Evidence for antiferromagnetic coupling of Fe layers across Cr interlayers, Phys. Rev. Lett. **57**, 2442 (1986).

[7] D. A. Allwood, G. Xiong, C. Faulkner, D. Atkinson, D. Petit, and R. Cowburn, Magnetic domain-wall logic, Science **309**, 1688 (2005).

[8] J. A. Currivan-Incorvia, S. Siddiqui, S. Dutta, E. R. Evarts, J. Zhang, D. Bono, C. Ross, and M. Baldo, Logic circuit prototypes for three-terminal magnetic tunnel junctions with mobile domain walls, Nat. Commun. **7**, 1 (2016).

[9] A. Imre, G. Csaba, L. Ji, A. Orlov, G. Bernstein, and W. Porod, Majority logic gate for magnetic quantum-dot cellular automata, Science **311**, 205 (2006).

[10] Z. Luo, A. Hrabec, T. P. Dao, G. Sala, S. Finizio, J. Feng, S. Mayr, J. Raabe, P. Gambardella, and L. J. Heyderman, Current-driven magnetic domain-wall logic, Nature **579**, 214 (2020).

[11] S. Li, W. Kang, Y. Huang, X. Zhang, Y. Zhou, and W. Zhao, Magnetic skyrmion-based artificial neuron device, Nanotechnology **28**, 31LT01 (2017).

[12] J. Grollier, D. Querlioz, K. Camsari, K. Everschor-Sitte, S. Fukami, and M. D. Stiles, Neuromorphic spintronics, Nat. Electron. **3**, 360 (2020).

[13] K. Yue, Y. Liu, R. K. Lake, and A. C. Parker, A brain-plausible neuromorphic on-the-fly learning system implemented with magnetic domain wall analog memristors, Sci. Adv. **5**, eaau8170 (2019).

[14] M. D. Stiles, Interlayer exchange coupling, J. Magn. Magn. Mater. **200**, 322 (1999).

[15] J. Nogués, J. Sort, V. Langlais, V. Skumryev, S. Suriñach, J. Muñoz, and M. Baró, Exchange bias in nanostructures, Phys. Rep. **422**, 65 (2005).

[16] E. E. Fullerton, J. Jiang, and S. Bader, Hard/soft magnetic heterostructures: model exchange-spring magnets, J. Magn. Magn. Mater. **200**, 392 (1999).

[17] L. Néel, Sur un nouveau mode de couplage entre les aimantations de deux couches minces ferromagnétiques, CR Acad. Sci **255**, 1676 (1962).

[18] B. Dieny, V. S. Speriosu, S. S. Parkin, B. A. Gurney, D. R. Wilhoit, and D. Mauri, Giant magnetoresistive in soft ferromagnetic multilayers, Phys. Rev. B **43**, 1297 (1991).

[19] N. Perrissin, S. Lequeux, N. Strelkov, A. Chavent, L. Vila, L. D. Buda-Prejbeanu, S. Auffret, R. C. Sousa, I. L. Prejbeanu, and B. Dieny, A highly thermally stable sub-20 nm magnetic random-access memory based on perpendicular shape anisotropy, Nanoscale **10**, 12187 (2018).

[20] G. Chen, A. Mascaraque, A. T. N'Diaye, and A. K. Schmid, Room temperature skyrmion ground state stabilized through interlayer exchange coupling, Appl. Phys. Lett. **106**, 242404 (2015).

[21] S.-H. Yang, K.-S. Ryu, and S. Parkin, Domain-wall velocities of up to 750 m s$^{-1}$ driven by exchange-coupling torque in synthetic antiferromagnets, Nat. Nanotech. **10**, 221 (2015).

[22] Y.-C. Lau, D. Betto, K. Rode, J. Coey, and P. Stamenov, Spin–orbit torque switching without an external field using interlayer exchange coupling, Nat. Nanotech. **11**, 758 (2016).

[23] A. van den Brink, G. Vermijs, A. Solignac, J. Koo, J. T. Kohlhepp, H. J. Swagten, and B. Koopmans, Field-free magnetization reversal by spin-Hall effect and exchange bias, Nat. Commun. **7**, 1 (2016).

[24] J. Yu, D. Bang, R. Mishra, R. Ramaswamy, J. H. Oh, H.-J. Park, Y. Jeong, P. Van Thach, D.-K. Lee, G. Go *et al.*, Long spin coherence length and bulk-like spin–orbit torque in ferrimagnetic multilayers, Nat. Mater. **18**, 29 (2019).

[25] V. Krizakova, K. Garello, E. Grimaldi, G. S. Kar, and P. Gambardella, Field-free switching of magnetic tunnel junctions driven by spin–orbit torques at sub-ns timescales, Appl. Phys. Lett. **116**, 232406 (2020).

[26] A. Hrabec, Z. Luo, L. J. Heyderman, and P. Gambardella, Synthetic chiral magnets promoted by the Dzyaloshinskii–Moriya interaction, Appl. Phys. Lett. **117**, 130503 (2020).

[27] S. H. Skjærvø, C. H. Marrows, R. L. Stamps, and L. J. Heyderman, Advances in artificial spin ice, Nat. Rev. Phys. **2**, 13 (2020).

[28] Z. Luo, T. P. Dao, A. Hrabec, J. Vijayakumar, A. Kleibert, M. Baumgartner, E. Kirk, J. Cui, T. Savchenko, G. Krishnaswamy *et al.*, Chirally coupled nanomagnets, Science **363**, 1435 (2019).

[29] T. P. Dao, M. Müller, Z. Luo, M. Baumgartner, A. Hrabec, L. J. Heyderman, and P. Gambardella, Chiral Domain Wall Injector Driven by Spin–Orbit Torques, Nano Lett. **19**, 5930 (2019).

[30] J. H. Franken, M. Herps, H. J. Swagten, and B. Koopmans, Tunable chiral spin texture in magnetic domain-walls, Sci. Rep. **4**, 1 (2014).

[31] Z. Luo, S. Schären, A. Hrabec, T. P. Dao, G. Sala, S. Finizio, J. Feng, S. Mayr, J. Raabe, P. Gambardella *et al.*, Field-and Current-Driven Magnetic Domain-Wall Inverter and Diode, Phys. Rev. Appl. **15**, 034077 (2021).

[32] Z. Zeng, Z. Luo, L. J. Heyderman, J.-V. Kim, and A. Hrabec, Synchronization of chiral vortex nano-oscillators, Appl. Phys. Lett. **118**, 222405 (2021).

[33] T. Devolder, J. Ferré, C. Chappert, H. Bernas, J.-P. Jamet, and V. Mathet, Magnetic properties of He+-irradiated Pt/Co/Pt ultrathin films, Phys. Rev. B **64**, 064415 (2001).

[34] M. Belmeguenai, J.-P. Adam, Y. Roussigné, S. Eimer, T. Devolder, J.-V. Kim, S. M. Cherif, A. Stashkevich, and A. Thiaville, Interfacial Dzyaloshinskii-Moriya interaction in perpendicularly magnetized Pt/Co/AlO x ultrathin films measured by Brillouin light spectroscopy, Phys. Rev. B **91**, 180405 (2015).

[35] E. Jué, A. Thiaville, S. Pizzini, J. Miltat, J. Sampaio, L. Buda-Prejbeanu, S. Rohart, J. Vogel, M. Bonfim, O. Boulle *et al.*, Domain wall dynamics in ultrathin Pt/Co/AlOx microstrips under large combined magnetic fields, Phys. Rev. B **93**, 014403 (2016).

[36] A. Thiaville, S. Rohart, É. Jué, V. Cros, and A. Fert, Dynamics of Dzyaloshinskii domain walls in ultrathin magnetic films, EPL (Europhysics Letters) **100**, 57002 (2012).

[37] S. Rohart and A. Thiaville, Skyrmion confinement in ultrathin film nanostructures in the presence of Dzyaloshinskii-Moriya interaction, Phys. Rev. B **88**, 184422 (2013).

[38] See Supplemental Material at [URL] for film deposition; device fabrication; characterization of Pt(5nm)|CoB(4.8nm)|Al(2nm); stopping field in CoB; comparison of Co and CoB system with/without dipolar coupling; and MuMax3 simulation parameters in CoB system

[39] J.-P. Tetienne, T. Hingant, L. Martínez, S. Rohart, A. Thiaville, L. H. Diez, K. Garcia, J.-P. Adam, J.-V. Kim, J.-F. Roch *et al.*, The nature of domain walls in ultrathin ferromagnets revealed by scanning nanomagnetometry, Nat. Commun. **6**, 1 (2015).

[40] S. Emori, U. Bauer, S.-M. Ahn, E. Martinez, and G. S. Beach, Current-driven dynamics of chiral ferromagnetic domain walls, Nat. Mater. **12**, 611 (2013).

[41] K.-S. Ryu, L. Thomas, S.-H. Yang, and S. Parkin, Chiral spin torque at magnetic domain walls, Nat. Nanotech. **8**, 527 (2013).

[42] A. Hrabec, N. Porter, A. Wells, M. Benitez, G. Burnell, S. McVitie, D. McGrouther, T. Moore, and C. Marrows, Measuring and tailoring the Dzyaloshinskii-Moriya interaction in perpendicularly magnetized thin films, Phys. Rev. B **90**, 020402 (2014).

[43] P. Metaxas, J. Jamet, A. Mougin, M. Cormier, J. Ferré, V. Baltz, B. Rodmacq, B. Dieny, and R. Stamps, Creep and flow regimes of magnetic domain-wall motion in ultrathin Pt/Co/Pt films with perpendicular anisotropy, Phys. Rev. Lett. **99**, 217208 (2007).

[44] I. M. Miron, G. Gaudin, S. Auffret, B. Rodmacq, A. Schuhl, S. Pizzini, J. Vogel, and P. Gambardella, Current-driven spin torque induced by the Rashba effect in a ferromagnetic metal layer, Nat. Mater. **9**, 230 (2010).

[45] E. Jué, C. Safeer, M. Drouard, A. Lopez, P. Balint, L. Buda-Prejbeanu, O. Boulle, S. Auffret, A. Schuhl, A. Manchon *et al.*, Chiral damping of magnetic domain walls, Nat. Mater. **15**, 272 (2016).

[46] R. Lavrijsen, G. Malinowski, J. Franken, J. Kohlhepp, H. Swagten, B. Koopmans, M. Czapkiewicz, and T. Stobiecki, Reduced domain wall pinning in ultrathin Pt/Co 100− x B x/Pt with perpendicular magnetic anisotropy, Appl. Phys. Lett. **96**, 022501 (2010).

[47] R. Lavrijsen, M. Verheijen, B. Barcones, J. Kohlhepp, H. Swagten, and B. Koopmans, Enhanced field-driven domain-wall motion in Pt/Co 68 B 32/Pt strips, Appl. Phys. Lett. **98**, 132502 (2011).

[48] F. Büttner, C. Moutafis, A. Bisig, P. Wohlhüter, C. M. Günther, J. Mohanty, J. Geilhufe, M. Schneider, C. v. K. Schmising, S. Schaffert *et al.*, Magnetic states in low-pinning high-anisotropy material nanostructures suitable for dynamic imaging, Phys. Rev. B **87**, 134422 (2013).

[49] S. Fukami, T. Suzuki, Y. Nakatani, N. Ishiwata, M. Yamanouchi, S. Ikeda, N. Kasai, and H. Ohno, Current-induced domain wall motion in perpendicularly magnetized CoFeB nanowire, Appl. Phys. Lett. **98**, 082504 (2011).

[50] Y. Yoshimura, K.-J. Kim, T. Taniguchi, T. Tono, K. Ueda, R. Hiramatsu, T. Moriyama, K. Yamada, Y. Nakatani, and T. Ono, Soliton-like magnetic domain wall motion induced by the interfacial Dzyaloshinskii–Moriya interaction, Nat. Phys. **12**, 157 (2016).

[51] A. Mougin, M. Cormier, J. Adam, P. Metaxas, and J. Ferré, Domain wall mobility, stability and Walker breakdown in magnetic nanowires, EPL (Europhysics Letters) **78**, 57007 (2007).

[52] S. A. Siddiqui, S. Dutta, A. Tang, L. Liu, C. A. Ross, and M. A. Baldo, Magnetic domain wall based synaptic and activation function generator for neuromorphic accelerators, Nano Lett. **20**, 1033 (2019).

[53] C. Cui, O. G. Akinola, N. Hassan, C. H. Bennett, M. J. Marinella, J. S. Friedman, and J. A. C. Incorvia, Maximized lateral inhibition in paired magnetic domain wall racetracks for neuromorphic computing, Nanotechnology **31**, 294001 (2020).

[54] T. Jin, W. Gan, F. Tan, N. Sernicola, W. S. Lew, and S. Piramanayagam, Synaptic element for neuromorphic computing using a magnetic domain wall device with synthetic pinning sites, J. phys., D, Appl. phys. **52**, 445001 (2019).

[55] T. Shibata, T. Shinohara, T. Ashida, M. Ohta, K. Ito, S. Yamada, Y. Terasaki, and T. Sasaki, Linear and symmetric conductance response of magnetic domain wall type spin-memristor for analog neuromorphic computing, Appl. Phys. Express **13**, 043004 (2020).

[56] F. Buijnsters, Y. Ferreiros, A. Fasolino, and M. Katsnelson, Chirality-dependent transmission of spin waves through domain walls, Phys. Rev. Lett. **116**, 147204 (2016).

[57] K. Wagner, A. Kákay, K. Schultheiss, A. Henschke, T. Sebastian, and H. Schultheiss, Magnetic domain walls as reconfigurable spin-wave nanochannels, Nat. Nanotech. **11**, 432 (2016).

[58] S. J. Hämäläinen, M. Madami, H. Qin, G. Gubbiotti, and S. van Dijken, Control of spin-wave transmission by a programmable domain wall, Nat. Commun. **9**, 1 (2018).

[59] J. Han, P. Zhang, J. T. Hou, S. A. Siddiqui, and L. Liu, Mutual control of coherent spin waves and magnetic domain walls in a magnonic device, Science **366**, 1121 (2019).

[60] Z. Liu, Z. Luo, S. Rohart, L. J. Heyderman, P. Gambardella, and A. Hrabec, Dataset for Engineering of intrinsic chiral torques in magnetic thin films based on the Dzyaloshinskii-Moriya interaction Zenodo.

## Appendix: Numerical model of effective field induced by the chiral edges

The dynamics of a domain wall in the system can be obtained from the energy of ⊙⊗⊙ and ⊙⊙⊙ configurations. We consider a case where a central $w_{OOP}$ wide track with perpendicular magnetization is surrounded by two in-plane magnetized zones with width $w_{IP}$, which are in turn surrounded by perpendicular magnetized zones as depicted in Fig. A1. While the outer zones are saturated along the ⊙ direction, the central track hosts a domain wall, which travels along the track ($x$ direction). The outer and central zones are coupled through the in-plane zones by a chiral interaction and dipolar coupling (the dipolar field flux closure also favours the ⊙⊗⊙ state).

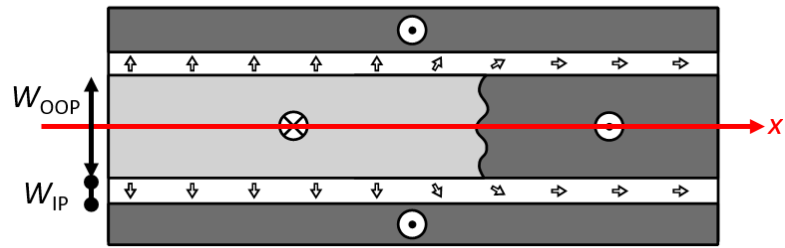

**FIG. A1**. Sketch of the sample hosting a domain wall in the central track.

The system energy depends on the DW position along $x$:

$$E(x) = E_{\odot\otimes\odot}x + (L - x)E_{\odot\odot\odot} + C_1, \quad (A1)$$

where $L$ is the track length and the constant $C_1$ contains DW position independent terms. $E_{\odot\otimes\odot}$ and $E_{\odot\odot\odot}$ correspond to the micromagnetic energy density integrated along the sample thickness and width ($z$ and $y$ directions) of the ⊙⊗⊙ and ⊙⊙⊙ states (i.e. far from the domain wall). Since the chiral and dipolar coupling favour the ⊙⊗⊙ state, $E_{\odot\otimes\odot} - E_{\odot\odot\odot} < 0$ and the energy is minimized when the domain wall moves along positive $x$. The domain wall experiences a force

$$F = -\frac{dE}{dx} = E_{\odot\odot\odot} - E_{\odot\otimes\odot}, \quad (A2)$$

oriented along positive $x$ direction.

This situation can be compared with the effect of a magnetic field in the absence of the chiral coupling. For a field $H$ oriented in the $\otimes$ direction, the system energy is

$$E_H(x) = 2\mu_0 H M_s w_{\text{OOP}} t\, x + C_2, \quad (A3)$$

where $t$ is the magnetic film thickness and $C_2$ is another constant independent of the domain wall position. Note that the factor 2 accounts for the $\pi$ magnetization rotation in the central track across the domain wall. The corresponding force is

$$F_H = -2\mu_0 H M_s w_{\text{OOP}} t. \quad (A4)$$

Comparing the two forces in equation A2 and A4, an effective field, oriented in the $\otimes$ direction, that describes the consequences of chiral edges, is determined as

$$H_{\text{eff}} = \frac{E_{\odot\otimes\odot} - E_{\odot\odot\odot}}{2 t \mu_0 M_s w_{\text{OOP}}}, \quad (A5)$$

This field is positive so it drives the domain wall along positive $x$. For the case where the coupling only arises from the Dzyaloshinskii-Moriya interaction, the coupling between the track and the outer zone is purely of the interfacial origin and $E_{\odot\otimes\odot} - E_{\odot\odot\odot}$ is independent on the track width, so the field is inversely proportional to $w_{\text{OOP}}$. If dipolar coupling is considered, the flux closure improves as the track width increases so $E_{\odot\otimes\odot} - E_{\odot\odot\odot}$ slightly increases with $w_{\text{OOP}}$ and the effective field decreases slower than $1/w_{\text{OOP}}$. The numerical estimation of the energy difference $E_{\odot\odot\odot} - E_{\odot\otimes\odot}$ in the Pt/CoB/AlOx trilayer, used to determine the effective fields in Fig. 5, are shown in Fig. A2.

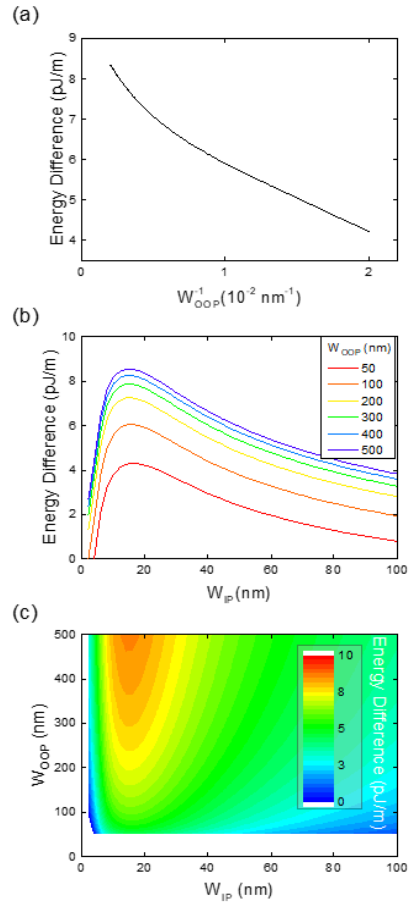

**FIG. A2**. (a) 1D micromagnetic calculation of energy difference $E_{\odot\odot\odot} - E_{\odot\otimes\odot}$ versus $1/w_{OOP}$ in a Pt/CoB/Al racetrack with fixed $w_{IP}$ = 20 nm. (b) Simulated energy difference as a function of $w_{IP}$ for $w_{OOP}$ = 50 nm up to 500 nm. (c) Color map of energy difference as a function of $w_{IP}$ and $w_{OOP}$.

**Supplementary Material:**

**Supplementary Note 1. Film deposition**

The multilayer magnetic film Pt(5nm)|Co(1.6nm)|Al(2nm) was deposited on a single side polished 525 μm-thick Si(100) substrate with 200 nm $Si_3N_4$ on the surface. The additional 200-nm-thick $Si_3N_4$ layer serves as an electrically insulating layer for transport measurements. The multilayer magnetic film Pt(5nm)|CoB(4.8nm)|Al(2nm) was deposited on single side polished 525-μm-thick Si(100) substrate. The composition of CoB is 68% Cobalt with 32% Boron. Both depositions are performed by DC magnetron sputtering at a base pressure below $3 \times 10^{-8}$ Torr. The Ar sputtering atmosphere pressure is 3 mTorr. The depositions are carried out with a commercial sputtering system from AJA International Inc.

**Supplementary Note 2. Device fabrication**

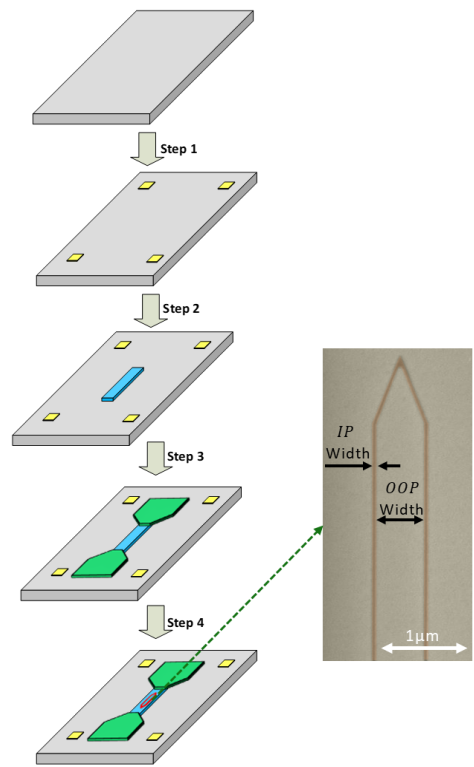

**FIG. S6**. Schematic of the nanofabrication process and false-color scanning electron micrograph of domain wall track device based on a Pt(5nm)|Co(1.6nm)|Al(2nm) trilayer. Here $w_{IP}$ = 55 nm and $w_{OOP}$ = 550 nm.

The schematic of the process flow for device fabrication is shown in Fig. S1 with the following fabrication steps:

1) Fabrication of gold markers for further pattern alignment. The gold markers are fabricated with electron beam lithography (E-beam resist: 4% percent PMMA with 950K molecular weight in ethyl lactate) followed by Cr(5nm)|Au(30nm)|Cr(5nm) metal evaporation at a base pressure of $2 \times 10^{-6}$ Torr and lift-off.
2) Magnetic film deposition (as described above) followed by lift-off.
3) Fabrication of Cr(5nm)|Au(30 nm)|Cr(5nm) electrodes with steps similar to those described in 1).
4) Patterning of oxidation mask: For an oxidation time of less than 40 s, a simple 2% 950K PMMA resist is used as an oxidation mask. For an oxidation time of more than 40 s, an additional metal evaporation and lift-off process is used to create a 10 nm-thick Cr layer as an oxidation mask. Oxygen plasma treatment is used to partially oxidize the unprotected aluminum layer.

**Supplementary Note 3. Characterization of Pt(5nm)|CoB(4.8nm)|Al(2nm)**

As shown in Fig. S2, by increasing the thickness of CoB layer from 4 nm to 5.2 nm, the magnetic anisotropy of the film can change from out-of-plane to in-plane anisotropy:

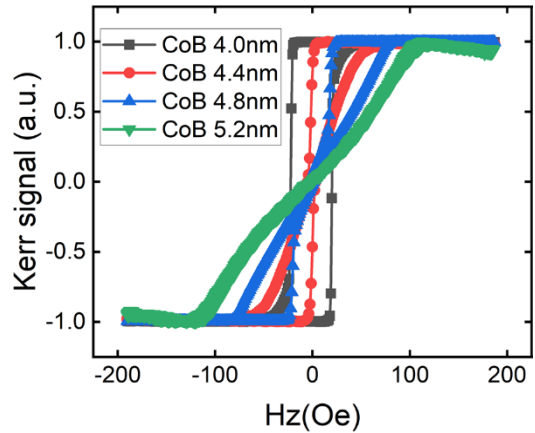

**FIG. S7**. Polar-MOKE measurement of Pt(5nm)|CoB(*x* nm)|Al(2nm) film, where *x* =4, 4.4, 4.8 and 5.2 nm. The measurement is in an out-of-plane magnetic field.

Thus, to maximize the effect of surface induced chiral-coupling, a CoB thickness of 4.8 nm is used as a starting point for the fabrication, since it is the thinnest film without magnetic remanence at zero external magnetic field.

By oxidizing of the top Al layer via oxygen plasma, the anisotropy of the film can change from in-plane anisotropy to out-of-plane anisotropy. In this work, we use a commercial Oxford RIE 100 machine to perform the oxygen plasma process. By applying an oxygen plasma at 30 W for 90 s, the magnetic anisotropy of the CoB layer switches from in-plane to out-of-plane as shown in Fig. S3.

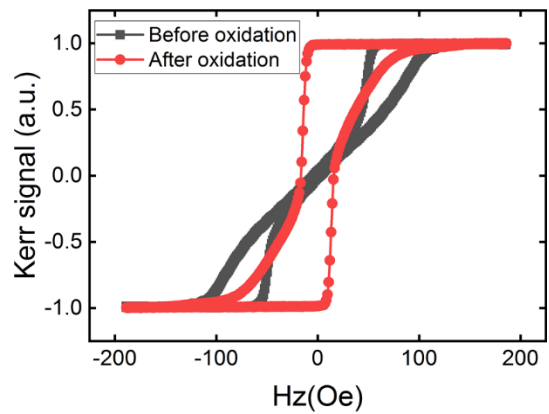

**FIG. S8**. Polar-MOKE measurement of Pt(5nm)|CoB(4.8nm)|Al(2nm) film before and after 30 W, 90 s oxygen plasma treatment.

To further understand the oxidation-induced out-of-plane anisotropy of the CoB sample, hysteresis loops are measured with a superconducting quantum interference device – vibrating sample magnetometer (SQUID-VSM) at room temperature in both an in-plane and out-of-plane external magnetic field (Fig. S4).

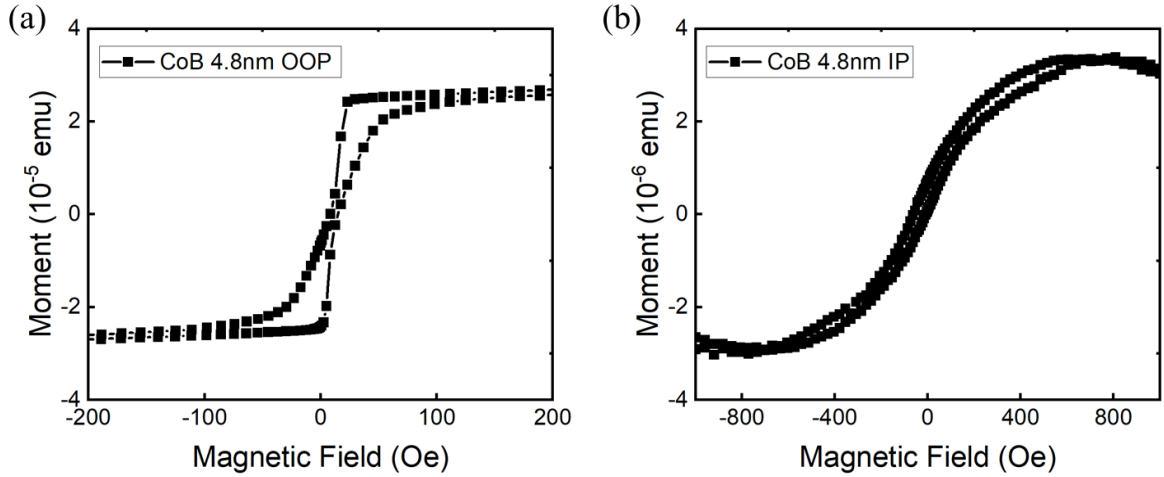

**FIG. S9**. Hysteresis loop measurement of Pt(5nm)|CoB(4.8nm)|AlOx(2nm) sample via SQUID under an a) out-of-plane magnetic field and b) in-plane magnetic field.

The saturation magnetization of the film $M_s = 4.01 \times 10^5$ A/m is extracted from the out-of-plane magnetic field measurement using:

$$M_s = \frac{m_s}{V}, \quad (S1)$$

where $m_s$ is the saturation magnetic moment and $V$ is the nominal volume of the CoB layer. The magnetic anisotropy constant $K_u = 1.14 \times 10^5$ J/m³ is determined using [49]

$$K_u = \frac{\mu_0 H_k M_s}{2}, \quad (S2)$$

where $H_k = H_{\text{saturation}} + M_s$.

### Supplementary Note 4. Stopping field in CoB

In a domain wall track device without chiral coupling, one needs to apply an external driving force (e.g. magnetic field or electrical current) to move a domain wall [50]. In our CoB system, which has low domain wall pinning, chiral coupling itself is able to overcome the pinning and thus displace the domain wall even in the absence of any external driving force (e.g. magnetic field or electric current). This is illustrated by micromagnetic simulations to calculate the domain wall velocity as shown in Fig. S5. In the absence of chiral coupling, the domain wall velocity increases when increasing the magnetic field until the so-called Walker breakdown field [51]. The velocities drop down sharply beyond the Walker field and then increase again. However, in a device with chiral coupling, the entire curve is shifted to the left due to the chiral coupling induced effective field. Thus, the zero velocity point is also left-shifted. We therefore define the stopping field, as the field at which the domain wall velocity is zero.

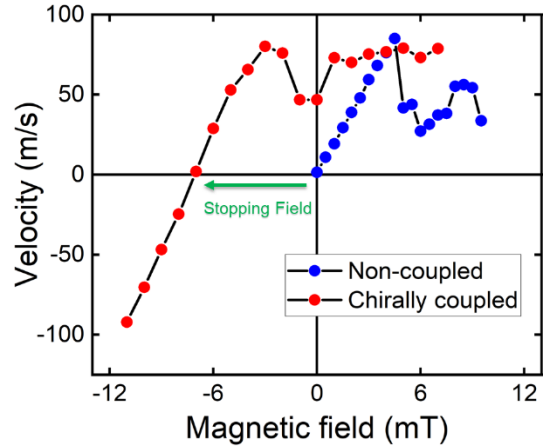

**FIG. S5**. Simulated domain wall velocity in a system with (red) and without (blue) chiral coupling. The stopping field magnitude is defined as the field at zero velocity as shown with the green arrow.

**Supplementary Note 5. Comparison of Co and CoB system with/without dipolar coupling**

To evaluate the influence of dipolar coupling due to the flux closure between the outer and inner part of DW track regions, 1D simulations for both Co and CoB system are performed. The material parameter used in these simulations are as follows: In the Pt/Co/AlOx stack, the Co thickness is 1.6 nm with $M_s = 9 \times 10^5$ A/m and interfacial DMI strength $D = 1.5 \times 10^{-3}$ J/m$^2$ [28]. For Pt/CoB/AlOx stack, the CoB thickness is 4 nm with $M_s = 4 \times 10^5$ A/m and interfacial DMI strength $D = 0.2 \times 10^{-3}$ J/m$^2$. A total width of 8192 nm is used and the width of the IP and OOP zones are varied. To extract the linear energy density of the ⊙⊗⊙ and ⊙⊙⊙ zones, two cells are used along the track with periodic boundary conditions.

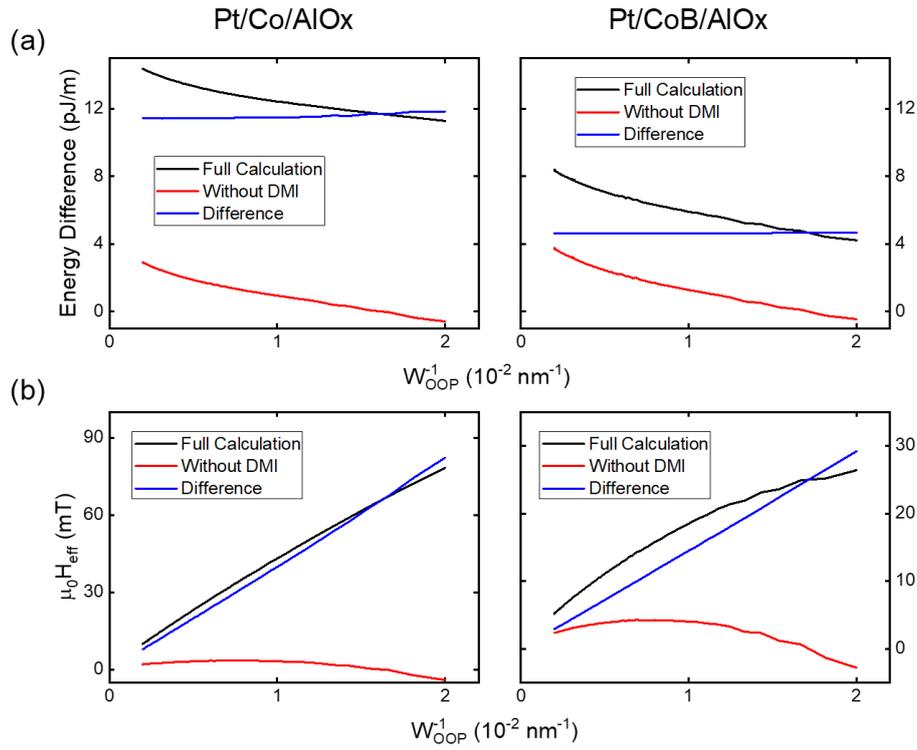

**FIG. S6**. (a) 1D micromagnetic simulation of energy difference ($E_{\odot\odot\odot} - E_{\odot\otimes\odot}$) versus $1/w_{OOP}$ in a racetrack with fixed $w_{IP}$ = 20 nm calculated in both Co and CoB based system. (b) $H_{eff}$ versus $1/w_{OOP}$ plots of both Co and CoB based system.

To reveal the impact of dipolar coupling, 1D calculations were performed in tracks with variable $w_{OOP}$ when DMI is absent and present, respectively. From the results displayed in Fig. S6(a), we can see that the effect of the dipolar coupling is equivalent in both system. However, a relative contribution to the total energy difference is more pronounced in the Pt/CoB/AlOx system where the DMI is weaker. Therefore, a linear $H_{eff}$ versus in $1/w_{OOP}$ dependence is observed in Pt/Co/AlOx system (Fig. S6 (b)) while a deviation from the linearity is observed in Pt/CoB/AlOx trilayer. This is because the DMI strength in the Co based system is almost one order of magnitude larger than in CoB system.

### Supplementary Note 6. MuMax3 simulation parameters in CoB system

The 2D stopping field simulation for CoB system is carried out using MuMax3. We define the geometry parameters as following: the IP width and OOP width of the track are varied, while the track length is fixed to 1600 nm. The width of OOP region outside the domain wall track is fixed to 1000 nm. For the material parameters of CoB, we set saturation magnetization to $4 \times 10^5$ A/m, the exchange stiffness to $A_{ex} = 5 \times 10^{-12}$ J/m and the Landau-Lifshitz damping constant to $\alpha = 0.1$. We varied interfacial DMI strength from 0.1 to $0.2 \times 10^{-3}$ J/m². The uniaxial anisotropy constant of the OOP region is fixed to $K_u(\text{OOP}) = 1.2 \times 10^5$ J/m³. Then two scenarios for the simulation are run with different IP properties, $K_u(\text{IP}) = 0$ for an ideal IP strip and $K_u(\text{IP}) = 0.8 K_u(\text{OOP})$, to include OOP anisotropy in the IP strip (only for 2D simulation).